\documentclass[prd,aps,onecolumn,preprintnumbers,amsmath,amssymb]{revtex4}
\usepackage{amsmath}
\usepackage{graphicx}
\usepackage{dcolumn}
\usepackage{bm}
\usepackage{amssymb}
\usepackage{latexsym}

\def\be{\begin{equation}}
\def\ee{\end{equation}}
\def\bea{\begin{eqnarray}}
\def\eea{\end{eqnarray}}

\begin{document}


\title{Non-Gaussianities of single field inflation with non-minimal coupling}

\author{Taotao Qiu\footnote{qiutt@mail.ihep.ac.cn} and Kwei-Chou Yang}

\affiliation{Department of Physics, Chung-Yuan Christian University,
Chung-li 320, Taiwan}
\begin{abstract}

We investigate the non-Gaussianities of inflation driven by a single
scalar field coupling non-minimally to the Einstein Gravity. We
assume that the form of the scalar field is very general with an
arbitrary sound speed. For convenience to study, we take the
subclass that the non-minimal coupling term is linear to the Ricci
scalar $R$. We define a parameter
$\mu\equiv\epsilon_h/\epsilon_\theta$ where $\epsilon_h$ and
$\epsilon_\theta$ are two kinds of slow-roll parameters, and obtain
the dependence of the shape of the 3-point correlation function on
$\mu$. We also show the estimator $F_{NL}$ in the equilateral limit.
Finally, based on numerical calculations, we present the
non-Gaussianities of non-minimal coupling chaotic inflation as an
explicit example.

\end{abstract}
\maketitle

\section{introduction}
The inflation Theory is one of the most successful theories of
modern cosmology. Having a period of very rapidly accelerating
expansion, it can not only solve many theoretical problems in
cosmology, such as flatness, horizon, monopole and so on, but also
gives the right amount of primordial fluctuations with nearly
scale-invariant power spectrum, which fits the data very well in
structure formation \cite{Guth:1980zm,Albrecht:1982wi,Linde:1983gd}.

There are many ways to construct inflation models, one of which is
to introduce a scalar field called ``inflaton" $\phi$ (see
\cite{Albrecht:1982wi,Linde:1983gd}). Moreover, one may expect that
inflaton could have non-minimal coupling to Ricci scalar $R$. The
most usual coupling form is $R\phi^2$, which was initially studied
for new inflation scenario \cite{Abbott:1981rg} and chaotic
inflation scenario \cite{Futamase:1987ua}. Later on, various models
have been taken on with deeply and wildly investigations. With a
non-minimal coupling term, inflation can be easily obtained and an
attractor solution is also available \cite{Futamase:1987ua}.
Perturbations based on non-minimal coupling inflation are discussed
in \cite{Salopek:1988qh}, where the coupling term may give rise to
corrections on power spectrum which can be used to fit the data or
constrain the parameters. Non-minimal couplings can be extended to
multifields, see \cite{Tsujikawa:2000tm}, or kinetic term coupling
\cite{Park:2008qf}. The constraints from observational data were
also performed, e.g. in \cite{Komatsu:1997hv}, where the authors
claimed that for non-minimal coupling chaotic inflation models, a
tiny tensor to scalar ratio will be obtained. Other applications of
non-minimal coupling inflation include the realization of warm
inflation \cite{Bellini:2002zr} and the avoidance of the so called
``$\eta$" problem \cite{Copeland:1994vg} in the framework of string
theory \cite{Easson:2009kk}. One can also see \cite{Faraoni:2000gx}
and also \cite{DeFelice:2010aj} for comprehensive reviews of
non-minimal coupling theories.

The non-Gaussianity of the primordial perturbation has been widely
acknowledged to be an important probe in the early universe
\cite{Bartolo:2004if,Lyth:2005fi,Boubekeur:2005fj,Komatsu:2009kd,
Chen:2010xka,Nitta:2009jp,Sefusatti:2009xu}. Experimentally, more
and more accurate data allow us to study the non-linear properties
of the fluctuation in Cosmic Microwave Background (CMB) and Large
Scale Structure (LSS) \cite{Gong:2009dt,Komatsu:2010fb,:2006uk};
Theoretically, the redundance of inflation models requires more
information than those of linear perturbations only to have them
distinguished. The non-Gaussianity of the fluctuations was first
considered in \cite{Allen:1987vq}, and it was further shown in
\cite{Acquaviva:2002ud} that the canonical single field slow roll
inflation can only give rise to negligible amount of
non-Gaussianity. To get large non-Gaussianity people need to find
new inflation models, an INCOMPLETE list and references of which
include: multi-field models \cite{Linde:1996gt}, k-inflation
\cite{Creminelli:2003iq}, DBI-type inflation
\cite{Alishahiha:2004eh}, curvaton scenario \cite{Lyth:2002my},
ghost inflation \cite{ArkaniHamed:2003uz}, warm inflation
\cite{Gupta:2002kn}, non-Bunch-Davies vacuum scenario
\cite{Holman:2007na}, bounce scenario \cite{Cai:2008ed}, island
cosmology \cite{Piao:2008dz}, loop correction \cite{Seery:2007we},
non-commutativity \cite{Fang:2007ba}, string gas scenario
\cite{Chen:2007js}, cosmic string \cite{Hindmarsh:2009qk},
``end-in-inflation" scenario \cite{Lyth:2005qk}, Ekpyrotic scenario
\cite{Koyama:2007if}, vector field \cite{Karciauskas:2008bc},
Ho$\breve{r}$ava theories \cite{Volovich:2009yh} and so on and so
forth\footnote{There are some other new mechanisms to generate large
non-Gaussianity, such as \cite{Dvali:2003em}.}.

In this note, we investigate the non-Gaussianity of inflation driven
by a general single field $P(X,\phi)$ coupling non-minimally to the
Einstein Gravity. Some specific examples of non-Gaussianities of
non-minimal coupled field has been studied in, e.g.,
\cite{Koh:2005ne} and non-Gaussianity generated by modified gravity
is expected to have effects that can be tested by CMB anisotropies
\cite{Gao:2010um}. By taking a subclass of linear coupling, we
calculated various shapes depending on the ratio between two slow
roll parameters $\epsilon_h$ and $\epsilon_\theta$, which describe
the evolution of cosmic expansion and the non-minimal correction,
respectively. The power spectrum will deviate from scale-invariance
due to the existence of non-minimal coupling \cite{DeFelice:2010aj},
and the shape of the 3-point correlation function are
correspondingly affected. In this paper we find that for different
(red or blue) tilt of the power spectrum, the shape will include
different parts which will obtain different amplitude of
non-Gaussianities. However, since we have only calculated up to
leading order in the slow-roll parameter, this conclusion has not
been so unambiguous yet. Nevertheless, If it can be verified after a
complete consideration to all the orders, one can find the relations
between 2- and 3-point correlation functions which can be used to
constrain non-minimal coupling models. This will be one of our
future works.

This paper is organized as follows: Sec.II briefly reviews the
preliminaries and basic equations of the general non-minimal
coupling single field inflation. We study the non-Gaussianities of
the general non-scalar field with linear coupling in Sec. III, which
is the main part of the paper. We first study the perturbed action
of the system up to 3rd order, and obtained the mode solution at the
quadratic level. After that, we calculate various shapes of the
3-point correlation functions using the mode solution. We also study
their equilateral limit and the relation with slow-roll parameters
at their leading order. In the last part of this section, we present
the non-Gaussianities of non-minimal coupling chaotic inflation as
an explicit example using numerical calculations. Sec. IV is the
conclusion and discussions.

\section{preliminary}
To begin with, let's consider the most general action of a single
scalar field with non-minimal coupling: \be\label{action}
S=\frac{1}{2}\int dtd^{3}x\sqrt{-g}[f(R,\phi)+2P(X,\phi)]~,\ee where
$X\equiv-\frac{1}{2}g^{\mu\nu}\partial_{\mu}\phi\partial_{\nu}\phi$
is the kinetic term and the metric
$g_{\mu\nu}=diag[-1,a^2(t),a^2(t),a^2(t)]$ with $a(t)$ the scale
factor of the universe. For the background evolution, one can vary
the action (\ref{action}) with respect to the field $\phi$ and the
metric $g_{\mu\nu}$ to get the equation of motion for $\phi$:
\be\label{eom} f_{\phi}+2P_{\phi}+2(P_{XX}\nabla^{\mu}X
+P_{X\phi}\nabla^{\mu}\phi)\nabla_{\mu}\phi+2P_{X}\Box\phi=0~,\ee
and the Einstein Equations: \be
\Sigma_{\mu\nu}=T_{\mu\nu}^{(\phi)}~,\ee where \bea
\Sigma_{\mu\nu}&\equiv&\Box
f_{R}g_{\mu\nu}-\nabla_{\mu}\nabla_{\nu}f_{R}+f_{R}R_{\mu\nu}-\frac{1}{2}fg_{\mu\nu}~,\\
T_{\mu\nu}^{(\phi)}&\equiv&P_{X}\nabla_{\mu}\phi\nabla_{\nu}\phi+Pg_{\mu\nu}~,\eea
with $\nabla_\mu$ being the covariant derivative with respect to the
metric $g_{\mu\nu}$ and $\Box\equiv\nabla_\mu\nabla^\mu$.

The evolution of the universe can be described by the slow roll
parameter: \be\label{epsilonh} \epsilon_h\equiv-\frac{\dot
H}{H^2}\ee where $H=\dot a/a$ is the Hubble parameter and dot means
the time derivative. In the inflation case, we require that
$\epsilon_h$ be small. Moreover, we can define two more parameters
following
\cite{Acquaviva:2002ud}:\bea \Sigma&\equiv&XP_X+2X^2P_{XX}~,\\
\lambda&\equiv&X^2P_{XX}+\frac{2}{3}X^3P_{XXX}~,\eea which will be
used in later parts of the paper.

To study non-Gaussianities, we adopt the usual convention of using
the {\it Arnowitt-Deser-Misner} (ADM) metric \cite{Arnowitt:1962hi}
as follows: \be\label{metric}
ds^{2}=-N^{2}dt^{2}+h_{ij}(dx^{i}+N^{i}dt)(dx^{j}+N^{j}dt)~,\ee
where $N(x)$ and $N_i(x)$ are the lapse function and the shift
vector, respectively. It is useful to decompose the action
(\ref{action}) into the 3+1 form, saying: \bea\label{actionadm}
S&=&\frac{1}{2}\int
dtd^{3}x\sqrt{h}N\{f[R^{(3)}+K_{ij}K^{ij}-K^{2}+\frac{2}{N\sqrt{h}}\partial_{t}(\sqrt{h}K)-\frac{2}{N\sqrt{h}}\partial_{i}(\sqrt{h}KN^{i}+\sqrt{h}h^{ij}\partial_{j}N),\phi]\nonumber\\
&&+2P(X,\phi)\}~.\eea The extrinsic tensor $K_{ij}$ is defined as:
\be
K_{ij}\equiv\frac{1}{2N}(\dot{h}_{ij}-\nabla_{i}N_{j}-\nabla_{j}N_{i})~\ee
where $\nabla_i$ is the covariant derivative with respect to the
metric $h_{ij}$ and its indices can be raised and lowered by
$h_{ij}$. The contraction $K\equiv K^i_i$. The three-dimensional
Ricci scalar $R^{(3)}$ is computed from the metric $h_{ij}$. From
this action, we are able to obtain the equations of motion for $N$
and $N_i$ (constraint equations) as: \bea\label{constraint1}
&&f-2f_{R}[K_{ij}K^{ij}-K^{2}+\frac{1}{\sqrt{h}N}\partial_{t}(\sqrt{h}K)-\frac{1}{\sqrt{h}N}\partial_{i}(\sqrt{h}KN^{i}+\sqrt{h}h^{ij}\partial_{j}N)]+\frac{2K}{N}(\partial_{t}f_{R}-N^{i}\partial_{i}f_{R})\nonumber\\
&&-\frac{2}{\sqrt{h}}\partial_{j}(\sqrt{h}h^{ij}\partial_{i}f_{R})+2P-2P_{X}v^{2}N^{-2}=0~,\eea
\be\label{constraint2}
\nabla_{j}(f_{R}K^{ij})-h^{ij}\nabla_{j}(f_{R}K)-h^{ij}\nabla_{j}(N^{-1}\partial_{t}f_{R})+h^{ij}\nabla_{j}(N^{-1}N^{l}\partial_{l}f_{R})+h^{ij}\partial_{j}f_{R}K-a^{-2}N^{-1}P_{X}v\partial^{i}\phi=0~,\ee
respectively, where $v\equiv\dot{\phi}-N^{i}\partial_{i}\phi$.
Moreover, a gauge choice is needed to eliminate the redundant
degrees of freedom. Here we choose the uniform density (comoving)
gauge, where the perturbations of the scalar field and metric take
the following form: \be\label{gauge}
\delta\phi=0~,~~~h_{ij}=a^{2}e^{2\zeta}\delta_{ij}~.\ee We also need
to expand the constraining variables $N$ and $N_i$. It is only
needed to expand them up to the $(n-2)$-th order when we calculate
$n$-th order perturbation \cite{Acquaviva:2002ud}. So here we expand
them to 1st order as follows: \be\label{expansion}
N=1+\alpha~,~~~N_{i}=\tilde{N}_{i}+\partial_{i}\psi~,~~~\partial^{i}\tilde{N}_{i}=0~,\ee
where $\alpha$, $\tilde{N}_{i}$ and $\psi$ are in the first order of
$\epsilon$.

\section{The non-Gaussianity calculation}

Since it is very complicated to consider the full types of
non-minimal coupling theory, in this paper we will only pick up a
subclass where the Ricci scalar is coupled linearly to the scalar
field, saying $f(R,\phi)=Rf_R(\phi)$. This type of non-minimal
coupling is often referred to as Scalar-Tensor Theory \cite{Fujii}
or linear coupling \cite{Qiu:2010ch}. For the non-linear coupling
case, the constraint equations are listed in {\bf Appendix A} and
the solutions will be postponed to the future studies.
\subsection{Up to 3rd Order Action}
Considering all the equations from (\ref{constraint1}) to
(\ref{expansion}) in the linear coupling case, one can get:\bea
&&-\frac{1}{2}a^{-2}f_{R0}\partial^{2}\tilde{N}^{i}+a^{-2}(2f_{R0}H+\dot{f_{R0}})\partial^{i}\alpha-2a^{-2}f_{R0}\partial^{i}\dot{\zeta}=0~,\\
&&4\alpha\Sigma+12Hf_{R0}\dot{\zeta}-12H^{2}f_{R0}\alpha-4a^{-2}f_{R0}H\partial^{2}\psi-4a^{-2}f_{R0}\partial^{2}\zeta-(12\alpha
H-6\dot{\zeta}+2a^{-2}\partial^{2}\psi)\dot{f_{R0}}=0~,\eea where
$f_{R0}$ is the background value of $f_R$ and for the current case
$f_{R0}=f_R$. The equations above give the specific solution at
first order in $\zeta$: \bea
\tilde{N}^{i}&=&0~,~~~\alpha=\frac{2f_{R0}\dot{\zeta}}{2f_{R0}H+\dot{f_{R0}}}~,\nonumber\\
\psi&=&\frac{-2f_{R0}}{2f_{R0}H+2\dot{f_{R0}}}\zeta+a^{2}\frac{3\dot{f_{R0}}^{2}+4f_{R0}\Sigma}{(2f_{R0}H+\dot{f_{R0}})^{2}}\partial^{-2}\dot{\zeta}~.\eea

We define $\theta\equiv\frac{1}{2}\ln(f_{R0}a^{2})$, so it will be
followed that $e^\theta=f_{R0}^\frac{1}{2}a$ and
$\dot\theta=H+\frac{\dot{f_{R0}}}{2f_{R0}}$. Note that when
$f_{R0}\rightarrow 1$, the system will return to the minimal
coupling case and $\dot\theta$ coincides with the Hubble parameter.
It is also convenient to rewrite $\alpha$ and $\psi$ in terms of
$\theta$ as:\be\label{solution}
\alpha=\frac{\dot{\zeta}}{\dot{\theta}}~,
~~~\psi=-\frac{\zeta}{\dot{\theta}}+\chi~,
~~~\chi=[3a^{2}(1-\frac{H}{\dot{\theta}})^{2}+a^{4}\dot{\theta}^{-2}e^{-2\theta}\Sigma]\partial^{-2}\dot{\zeta}~.\ee
Hereafter, we will use this parameter for convenience throughout the
paper.

Expanding the action (\ref{actionadm}) to 3rd order of $\zeta$ and
substituting Eq. (\ref{solution}) into the action, one can get the
expanded action in each order: \be\label{s0} S_{0}=\frac{1}{2}\int
dtd^{3}xa^{3}(f_{0}+2P_{0})~,\ee \be\label{s1} S_{1}=0~,\ee
\be\label{s2} S_{2}=\int
dtd^{3}xae^{2\theta}\{3(\frac{H^{2}}{\dot{\theta}^{2}}-2\frac{H}{\dot{\theta}}+1)\dot{\zeta}^{2}+a^{2}e^{-2\theta}\Sigma\frac{\dot{\zeta}^{2}}{\dot{\theta}^{2}}+a^{-2}(\frac{H}{\dot{\theta}}-1+\frac{\ddot{\theta}}{\dot{\theta}^{2}})(\partial\zeta)^{2}\}~,\ee
\bea\label{s3} S_{3}&=&\frac{1}{2}\int
dtd^{3}xae^{2\theta}\{+6(+2\frac{H}{\dot{\theta}}-1-\frac{H^{2}}{\dot{\theta}^{2}})\frac{\dot{\zeta}^{3}}{\dot{\theta}}+18(-2\frac{H}{\dot{\theta}}+1+\frac{H^{2}}{\dot{\theta}^{2}})\zeta\dot{\zeta}^{2}+2a^{-2}(\frac{H}{\dot{\theta}}-1)\zeta(\partial\zeta)^{2}\nonumber\\
&&+2a^{-2}\frac{\ddot{\theta}\zeta}{\dot{\theta}^{2}}(\partial\zeta)^{2}-4a^{-4}(\partial\zeta\cdot\partial\psi)\partial^{2}\psi+3a^{-4}\zeta\partial_{i}\partial_{j}\psi\partial^{j}\partial^{i}\psi-a^{-4}\frac{\dot{\zeta}}{\dot{\theta}}\partial_{i}\partial_{j}\psi\partial^{i}\partial^{j}\psi-3a^{-4}\zeta(\partial^{2}\psi)^{2}\nonumber\\
&&+a^{-4}\frac{\dot{\zeta}}{\dot{\theta}}(\partial^{2}\psi)^{2}\}+\frac{1}{2}\int
dtd^{3}xa^{3}\{6\Sigma\frac{\dot{\zeta}^{2}}{\dot{\theta}^{2}}\zeta-2(\Sigma+2\lambda)\frac{\dot{\zeta}^{3}}{\dot{\theta}^{3}}\}~.\eea

This process is very straightforward, but rather tedious. The
physical meaning of each equation is easy understanding: The 0th
order expansion (\ref{s0}) is just the background part of the
original action (\ref{action}) where subscript `0' denotes the
background value; the 1st order expansion (\ref{s1}) is just the
background equation of motion. The 2nd and 3rd order expansions are
only deviate from the GR case due to the difference of the parameter
$\dot\theta$ from $H$, which will coincide when $f_{R0}\rightarrow
1$. However, as will be seen below, this deviation causes very
different results of non-Gaussianity of our case from that of GR.
\subsection{Quadratic Part: Mode Solution}
First of all, let's consider the solution of the 2nd order action
(\ref{s2}). This is the most important and kernel step in the
calculation of the bispectrum which will be performed later. The 2nd
order action can be written as: \bea S_{2}&=&\int
dtd^{3}xae^{2\theta}\{3(\frac{H^{2}}{\dot{\theta}^{2}}-2\frac{H}{\dot{\theta}}+1)\dot{\zeta}^{2}+a^{2}e^{-2\theta}\Sigma\frac{\dot{\zeta}^{2}}{\dot{\theta}^{2}}+a^{-2}(\frac{H}{\dot{\theta}}-1+\frac{\ddot{\theta}}{\dot{\theta}^{2}})(\partial\zeta)^{2}\}\nonumber\\
&=&\int d\tau
d^{3}xe^{2\theta}\{3(\frac{H^{2}}{\dot{\theta}^{2}}-2\frac{H}{\dot{\theta}}+1){u'_{\overrightarrow{k}}}^{2}(\tau)+a^{2}e^{-2\theta}\frac{\Sigma}{\dot{\theta}^{2}}{u'_{\overrightarrow{k}}}^{2}(\tau)-(\frac{H}{\dot{\theta}}-1+\frac{\ddot{\theta}}{\dot{\theta}^{2}})k^2u_{\overrightarrow{k}}^{2}(\tau)\}~,\eea
where in the second step we have used conformal time $\tau\equiv\int
\frac{dt}{a(t)}$ and transformed the variable
$\zeta(\tau,\overrightarrow{x})$ into its Fourier form, namely: \be
\zeta(\tau,\overrightarrow{x})=\int\frac{d\overrightarrow{k}}{\sqrt{2E}}[u_{\overrightarrow{k}}(\tau)a_{\overrightarrow{k}}e^{-i\overrightarrow{k}\cdot\overrightarrow{x}}+u_{\overrightarrow{k}}^{\ast}(\tau)a_{\overrightarrow{k}}^{\dagger}e^{i\overrightarrow{k}\cdot\overrightarrow{x}}]~.\ee

It is convenient to define another variable $v_k$ as $v_k\equiv
zu_{\overrightarrow{k}}(\tau)$, where \be
z=\sqrt{\Biggl|3e^{2\theta}(\frac{H}{\dot{\theta}}-1)^{2}+\frac{a^{2}\Sigma}{\dot{\theta}^{2}}\Biggl|}~,\ee
to let the equation of motion for $v_k$ be in a canonical form,
which is: \be\label{vk}
v_{k}^{\prime\prime}+(c_s^2k^{2}-\frac{z^{\prime\prime}}{z})v_{k}=0~.\ee
Here the effective sound speed squared $c_s^2$ is defined as
$c_s^2=(\dot{\theta}^{2}-H\dot{\theta}-\ddot{\theta})/\biggl|3(H-\dot{\theta})^{2}+f_{R0}^{-1}\Sigma\biggl|$,
and thus one can have
$z=\frac{e^{\theta}}{c_{s}}\sqrt{\epsilon_{\theta}+1-\frac{H}{\dot{\theta}}}$.

In the inflation period, the Hubble parameter changes slowly so as
to have enough fast expansion. Here in order to solve the equation
above, we need to introduce another parameter: \be
\epsilon_\theta\equiv-\frac{\ddot\theta}{\dot\theta^2}~,\ee which
describes the variation of $f_{R0}$ with respect to time. In case
that $\epsilon_h$ (see Eq. (\ref{epsilonh})) and $\epsilon_\theta$
are both small and $\dot\epsilon_i\ll\epsilon_i(i=h,\theta)$, we can
take the leading order of $\epsilon_h$ and $\epsilon_\theta$ so that
Eq. (\ref{vk}) becomes: \be\label{mode}
v_{\overrightarrow{k}}^{\prime\prime}+(c_{s}^{2}k^{2}-\frac{\mu^{2}+\mu}{\tau^{2}})v_{\overrightarrow{k}}=0~,\ee
where $\mu\equiv\epsilon_{h}/\epsilon_{\theta}$ is the ratio of the
two slow roll parameters. The solution of Eq. (\ref{mode}) can be
presented in form of Hankal function: \be\label{hankel}
v_{\overrightarrow{k}}(\tau)=C\sqrt{c_{s}k|\tau|}H_{\pm(\mu+\frac{1}{2})}(c_{s}k|\tau|)~,\ee
where $C$ is an undetermined constant denoting the amplitude of the
solution. In deriving this, the approximation of slow-varying sound
speed $\dot c_s<<1$ is also taken.

The solution (\ref{hankel}) can be splitted into two limits
corresponding to the subhubble and superhubble regions respectively.
In the superhubble region, we can take the limit to be:\bea
v_{\overrightarrow{k}}&\rightarrow&\sqrt{c_{s}k|\tau|}(\frac{C}{\Gamma(\mu+\frac{3}{2})}(c_{s}k|\tau|)^{\mu+\frac{1}{2}}+\frac{C}{\Gamma(-\mu+\frac{1}{2})}(c_{s}k|\tau|)^{-(\mu+\frac{1}{2})})\nonumber\\
\label{super1}&=&\frac{C}{\Gamma(\mu+\frac{3}{2})}(c_{s}k|\tau|)^{\mu+1}+\frac{C}{\Gamma(-\mu+\frac{1}{2})}(c_{s}k|\tau|)^{-\mu}~,\eea
and from the relation $v_k=zu_{\overrightarrow{k}}(\tau)$ one have:
\bea
u_{\overrightarrow{k}}&=&\frac{e^{-\theta}c_{s}}{\sqrt{|\epsilon_{\theta}+1-\frac{H}{\dot{\theta}}|}}\{\frac{C}{\Gamma(\mu+\frac{3}{2})}(c_{s}k|\tau|)^{\mu+1}+\frac{C}{\Gamma(-\mu+\frac{1}{2})}(c_{s}k|\tau|)^{-\mu}\}\nonumber\\
\label{super2}&=&\frac{\mu^{\frac{1}{2}}H}{f_{R0}^{\frac{1}{2}}(-1)c_{s}^{\mu-1}k^{\mu}|\epsilon_{h}+\mu-1|^{\frac{1}{2}}}\{\frac{C}{\Gamma(\mu+\frac{3}{2})}(c_{s}k|\tau|)^{2\mu+1}+\frac{C}{\Gamma(-\mu+\frac{1}{2})}\}\eea
where $f_{R0}(-1)$ denotes the value of $f_{R0}$ at $\tau=1$ and
$f_{R0}$ can be parameterized as $f_{R0}\simeq
f_{R0}(-1)|\tau|^{2(1-\mu)}$. In deriving these equations we also
used the approximation $\tau=-1/(aH)+{\cal{O}}(\epsilon_h)$. The
solution contains a constant mode and a decaying mode, the latter of
which is irrelevant and should be discarded. In the subhubble
region, we can take the limit as: \be\label{sub1}
v_{\overrightarrow{k}}
\rightarrow
C\sqrt{\frac{2}{\pi}}e^{ic_{s}k|\tau|}e^{i\frac{\mu\pi}{2}}~,\ee
where we also discarded the $+(\mu+\frac{1}{2})$ branch. On the
other hand, one can use WKB method to calculate the subhubble
solution of Eq. (\ref{mode}), which is: \be\label{sub2}
v_{\overrightarrow{k}}
\simeq\frac{i\mathcal{H}}{\sqrt{8c_{s}^{3}k^{3}}}e^{-ic_{s}k\tau}[\mu(1+\mu)+2ic_{s}k\tau]~.\ee
Comparing Eqs. (\ref{sub1}) and (\ref{sub2}) at
$\tau\rightarrow-\infty$, one can determine the coefficient $C$
as:\be\label{c}
C=\sqrt{\frac{\pi}{4c_{s}k}}e^{-i\frac{\mu\pi}{2}}~.\ee With this in
hand, we can have the exact solution of $u_{\overrightarrow{k}}$ in
both superhubble and subhubble limits. Substituting this back to
(\ref{super2}), we get the final form of the superhubble solution:
\be
u_{\overrightarrow{k}}=\frac{\mu^{\frac{1}{2}}H}{2f_{R0}^{\frac{1}{2}}(-1)c_{s}^{\mu-1}k^{\mu}|\epsilon_{h}+\mu-1|^{\frac{1}{2}}\Gamma(-\mu+\frac{1}{2})}\sqrt{\frac{\pi}{c_{s}k}}e^{-i\frac{\mu\pi}{2}}~.\ee
It is a time-independent mode and thus can be applied to far future
where $\tau=0$. From this we can also obtain the power spectrum of
$\zeta$, which is: \bea
\mathcal{P}_{k}^{\zeta}&\equiv&\frac{k^{3}}{2\pi^{2}}|u_{\overrightarrow{k}}|^{2}\nonumber\\
\label{spectrum}&=&\frac{\mu H^{2}}{8\pi
f_{R0}(-1)c_{s}^{2\mu-1}k^{2\mu-2}|\epsilon_{h}+\mu-1|\Gamma^{2}(-\mu+\frac{1}{2})}~,\eea
and the spectrum index is
$n_\zeta\equiv\frac{d\ln{\mathcal{P}_{k}^{\zeta}}}{d\ln{k}}+1=2(1-\mu)+1$.
One can see from this that the power gets a red spectrum
($n_\zeta<1$) when $\mu>1$ while a blue spectrum ($n_\zeta>1$) will
be obtained at $\mu<1$. Moreover, the constraints that the
primordial spectrum must be nearly scale invariant requires that
$|\mu-1|\sim{\cal O}(\epsilon)$.

Furthermore, using Eq. (\ref{sub1}), the subhubble solution can be
solved as: \bea u_{\overrightarrow{k}}
&\simeq&\frac{i\mathcal{H}e^{-\theta}\mu^{\frac{1}{2}}}{2\sqrt{2}c_{s}^{\frac{1}{2}}|\overrightarrow{k}|^{\frac{3}{2}}|\epsilon_{h}+\mu-1|^{\frac{1}{2}}}e^{-ic_{s}|\overrightarrow{k}|\tau}[\mu(1+\mu)+2ic_{s}|\overrightarrow{k}|\tau]\nonumber\\
&=&\frac{iH\mu^{\frac{1}{2}}}{2\sqrt{2}c_{s}^{\frac{1}{2}}k^{\frac{3}{2}}f_{R0}^{\frac{1}{2}}|\epsilon_{h}+\mu-1|^{\frac{1}{2}}}e^{-ic_{s}k\tau}[\mu(1+\mu)+2ic_{s}k\tau]\nonumber\\
&=&\frac{iH\mu^{\frac{1}{2}}}{2\sqrt{2}c_{s}^{\frac{1}{2}}k^{\frac{3}{2}}f_{R0}^{\frac{1}{2}}(-1)|\epsilon_{h}+\mu-1|^{\frac{1}{2}}}e^{-ic_{s}k\tau}[\mu(1+\mu)|\tau|^{\mu-1}-2ic_{s}k|\tau|^{\mu}]~,\eea
and \be
\frac{d}{d\tau}u_{\overrightarrow{k}}^{\ast}(\tau)=\frac{iH\mu^{\frac{1}{2}}c_{s}^{\frac{3}{2}}k^{\frac{1}{2}}e^{ic_{s}|\overrightarrow{k}|\tau}}{\sqrt{2}f_{R0}^{\frac{1}{2}}(-1)|\epsilon_{h}+\mu-1|^{\frac{1}{2}}}|\tau|^{\mu}~,\ee
where we keep only the leading order terms in terms of $\epsilon$.
The results above will be useful for our analysis of non-Gaussianity
in the next paragraph.

\subsection{Cubic Part: Non-Gaussianities}
According to ``in-in" formalism \cite{Weinberg:2005vy}, the 3-point
correlation function is characterized in the interaction picture as:
\be\label{inin}
<|\zeta(\tau,\overrightarrow{k_1})\zeta(\tau,\overrightarrow{k_2})\zeta(\tau,\overrightarrow{k_3})|>=-i\mathcal{T}\int_{t_{0}}^{t}dt^{\prime}<|[\zeta(t,\overrightarrow{k_{1}})\zeta(t,\overrightarrow{k_{2}})\zeta(t,\overrightarrow{k_{3}}),H_{int}^p(t^{\prime})]|>~,\ee
where $H_{int}^p$ is the 3rd order interaction Hamiltonian and
$\mathcal{T}$ is the time-ordering operator. From the 3rd order
action (\ref{s3}), we can write down the 3rd order Hamiltonian as:
\bea H_{int}&=&\int
dtd^{3}xae^{2\theta}\{[a^{2}e^{-2\theta}(\Sigma+2\lambda)+3(\dot{\theta}-H)^{2}]\frac{\dot{\zeta}^{3}}{\dot{\theta}^{3}}-3[a^{2}e^{-2\theta}\Sigma+3(\dot{\theta}-H)^{2}]\frac{\zeta\dot{\zeta}^{2}}{\dot{\theta}^{2}}-a^{-2}(\frac{H}{\dot{\theta}}-1+\frac{\ddot{\theta}}{\dot{\theta}^{2}})\zeta(\partial\zeta)^{2}\nonumber\\
&&-a^{-4}\dot{\theta}^{-1}[3a^{2}(1-\frac{H}{\dot{\theta}})^{2}+a^{4}\dot{\theta}^{-2}e^{-2\theta}\Sigma]\dot{\zeta}(\partial\zeta)^{2}+2a^{-4}[3a^{2}(1-\frac{H}{\dot{\theta}})^{2}+a^{4}\dot{\theta}^{-2}e^{-2\theta}\Sigma]\dot{\zeta}\partial\zeta\partial\chi\}~,\eea
or, if changed to momentum space, \bea
H_{int}^{p}&=&\int\frac{d^{3}p_{1}d^{3}p_{2}d^{3}p_{3}}{(2\pi)^{9}}(2\pi)^{3}\delta^{3}(\overrightarrow{p_{1}}+\overrightarrow{p_{2}}+\overrightarrow{p_{3}})ae^{2\theta}\{[a^{2}e^{-2\theta}(\Sigma+2\lambda)+3(\dot{\theta}-H)^{2}]\frac{\dot{\zeta}(t,\overrightarrow{p_{1}})\dot{\zeta}(t,\overrightarrow{p_{2}})\dot{\zeta}(t,\overrightarrow{p_{3}})}{\dot{\theta}^{3}}\nonumber\\
&&-3[a^{2}e^{-2\theta}\Sigma+3(\dot{\theta}-H)^{2}]\frac{\zeta(t,\overrightarrow{p_{1}})\dot{\zeta}(t,\overrightarrow{p_{2}})\dot{\zeta}(t,\overrightarrow{p_{3}})}{\dot{\theta}^{2}}-a^{-2}(\frac{H}{\dot{\theta}}-1+\frac{\ddot{\theta}}{\dot{\theta}^{2}})(\overrightarrow{p_{2}}\cdot\overrightarrow{p_{3}})\zeta(t,\overrightarrow{p_{1}})\zeta(t,\overrightarrow{p_{2}})\zeta(t,\overrightarrow{p_{3}})\nonumber\\
&&-a^{-4}\dot{\theta}^{-1}[3a^{2}(1-\frac{H}{\dot{\theta}})^{2}+a^{4}\dot{\theta}^{-2}e^{-2\theta}\Sigma](\overrightarrow{p_{2}}\cdot\overrightarrow{p_{3}})\dot{\zeta}(t,\overrightarrow{p_{1}})\zeta(t,\overrightarrow{p_{2}})\zeta(t,\overrightarrow{p_{3}})\nonumber\\
&&+2a^{-4}[3a^{2}(1-\frac{H}{\dot{\theta}})^{2}+a^{4}\dot{\theta}^{-2}e^{-2\theta}\Sigma](\overrightarrow{p_{2}}\cdot\overrightarrow{p_{3}})\dot{\zeta}(t,\overrightarrow{p_{1}})\zeta(t,\overrightarrow{p_{2}})\chi(t,\overrightarrow{p_{3}})\}~.\eea
There are five terms of 3rd order, each containing a long prefactor.
Using Eq. (\ref{inin}), we can calculate their contributions to
non-Gaussianities. Neglecting the detailed calculating process, we
only give the final results of the contributions to
non-Gaussianities from each term as follows.

The contribution from $\dot{\zeta}^{3}$: \bea
&&-6i\frac{\Sigma+2\lambda}{\mu^{3}H^{4}}u_{\overrightarrow{k_{1}}}(0)u_{\overrightarrow{k_{2}}}(0)u_{\overrightarrow{k_{3}}}(0)\int_{-\infty}^{0}\frac{-1}{\tau}d\tau(2\pi)^{3}\delta^{3}(\sum_{i}\overrightarrow{k_{i}})\frac{d}{d\tau}u_{-\overrightarrow{k_{1}}}^{\ast}(\tau)\frac{d}{d\tau}u_{-\overrightarrow{k_{2}}}^{\ast}(\tau)\frac{d}{d\tau}u_{-\overrightarrow{k_{3}}}^{\ast}(\tau)\nonumber\\
&&-18i\frac{(\mu-1)^{2}f_{R0}(-1)}{\mu^{3}H^{2}}u_{\overrightarrow{k_{1}}}(0)u_{\overrightarrow{k_{2}}}(0)u_{\overrightarrow{k_{3}}}(0)\int_{-\infty}^{0}|\tau|^{1-2\mu}d\tau(2\pi)^{3}\delta^{3}(\sum_{i}\overrightarrow{k_{i}})\frac{d}{d\tau}u_{-\overrightarrow{k_{1}}}^{\ast}(\tau)\frac{d}{d\tau}u_{-\overrightarrow{k_{2}}}^{\ast}(\tau)\frac{d}{d\tau}u_{-\overrightarrow{k_{3}}}^{\ast}(\tau)\nonumber\\
\label{con1}&&+c.c.,\eea The contribution from
$\zeta\dot{\zeta}^{2}$: \bea
&&6i\frac{\Sigma}{\mu^{2}H^{4}}u_{\overrightarrow{k_{1}}}(0)u_{\overrightarrow{k_{2}}}(0)u_{\overrightarrow{k_{3}}}(0)\int_{-\infty}^{0}\frac{1}{\tau^{2}}d\tau(2\pi)^{3}\delta^{3}(\sum_{i}\overrightarrow{k_{i}})[u_{-\overrightarrow{k_{1}}}^{\ast}(\tau)\frac{d}{d\tau}u_{-\overrightarrow{k_{2}}}^{\ast}(\tau)\frac{d}{d\tau}u_{-\overrightarrow{k_{3}}}^{\ast}(\tau)+2perms.]\nonumber\\
&&18i\frac{(\mu-1)^{2}f_{R0}(-1)}{\mu^{2}H^{2}}u_{\overrightarrow{k_{1}}}(0)u_{\overrightarrow{k_{2}}}(0)u_{\overrightarrow{k_{3}}}(0)\int_{-\infty}^{0}|\tau|^{-2\mu}d\tau(2\pi)^{3}\delta^{3}(\sum_{i}\overrightarrow{k_{i}})[u_{-\overrightarrow{k_{1}}}^{\ast}(\tau)\frac{d}{d\tau}u_{-\overrightarrow{k_{2}}}^{\ast}(\tau)\frac{d}{d\tau}u_{-\overrightarrow{k_{3}}}^{\ast}(\tau)+2perms.]\nonumber\\
\label{con2}&&+c.c.,~\eea The contribution from
$\zeta(\partial\zeta)^{2}$: \bea &&2i\frac{(\mu-1)f_{R0}(-1)}{\mu
H^{2}}u_{\overrightarrow{k_{1}}}(0)u_{\overrightarrow{k_{2}}}(0)u_{\overrightarrow{k_{3}}}(0)\int_{-\infty}^{0}|\tau|^{-2\mu}d\tau(2\pi)^{3}\delta^{3}(\sum_{i}\overrightarrow{k_{i}})[(\overrightarrow{k_{2}}\cdot\overrightarrow{k_{3}})u_{-\overrightarrow{k_{1}}}^{\ast}(\tau)u_{-\overrightarrow{k_{2}}}^{\ast}(\tau)u_{-\overrightarrow{k_{3}}}^{\ast}(\tau)\nonumber\\
\label{con3}&&+2perms.]+c.c.,\eea The contribution from
$\dot{\zeta}(\partial\zeta)^{2}$: \bea
&&-6i\frac{(\mu-1)^{2}f_{R0}(-1)}{\mu^{3}H^{2}}u_{\overrightarrow{k_{1}}}(0)u_{\overrightarrow{k_{2}}}(0)u_{\overrightarrow{k_{3}}}(0)\int_{-\infty}^{0}|\tau|^{1-2\mu}d\tau(2\pi)^{3}\delta^{3}(\sum_{i}\overrightarrow{k_{i}})[(\overrightarrow{k_{2}}\cdot\overrightarrow{k_{3}})\frac{d}{d\tau}u_{-\overrightarrow{k_{1}}}^{\ast}(\tau)u_{-\overrightarrow{k_{2}}}^{\ast}(\tau)u_{-\overrightarrow{k_{3}}}^{\ast}(\tau)+2perms]\nonumber\\
\label{con4}&&-2i\frac{\Sigma}{\mu^{3}H^{4}}u_{\overrightarrow{k_{1}}}(0)u_{\overrightarrow{k_{2}}}(0)u_{\overrightarrow{k_{3}}}(0)\int_{-\infty}^{0}\frac{-1}{\tau}d\tau(2\pi)^{3}\delta^{3}(\sum_{i}\overrightarrow{k_{i}})[(\overrightarrow{k_{2}}\cdot\overrightarrow{k_{3}})\frac{d}{d\tau}u_{-\overrightarrow{k_{1}}}^{\ast}(\tau)u_{-\overrightarrow{k_{2}}}^{\ast}(\tau)u_{-\overrightarrow{k_{3}}}^{\ast}(\tau)+2perms]+c.c.,\eea
The contribution from $\dot{\zeta}\partial\zeta\partial\chi$: \bea
&&-18i\frac{(\mu-1)^{4}f_{R0}(-1)}{\mu^{4}H^{2}}u_{\overrightarrow{k_{1}}}(0)u_{\overrightarrow{k_{2}}}(0)u_{\overrightarrow{k_{3}}}(0)\int_{-\infty}^{0}|\tau|^{-2\mu}d\tau(2\pi)^{3}\delta^{3}(\sum_{i}\overrightarrow{k_{i}})[\frac{\overrightarrow{k_{2}}\cdot\overrightarrow{k_{3}}}{k_{3}^{2}}\frac{d}{d\tau}u_{-\overrightarrow{k_{1}}}^{\ast}(\tau)u_{-\overrightarrow{k_{2}}}^{\ast}(\tau)\frac{d}{d\tau}u_{-\overrightarrow{k_{3}}}^{\ast}(\tau)+5perms]\nonumber\\
&&-2i\frac{\Sigma^{2}}{\mu^{4}H^{6}f_{R0}(-1)}u_{\overrightarrow{k_{1}}}(0)u_{\overrightarrow{k_{2}}}(0)u_{\overrightarrow{k_{3}}}(0)\int_{-\infty}^{0}|\tau|^{2\mu-4}d\tau(2\pi)^{3}\delta^{3}(\sum_{i}\overrightarrow{k_{i}})[\frac{\overrightarrow{k_{2}}\cdot\overrightarrow{k_{3}}}{k_{3}^{2}}\frac{d}{d\tau}u_{-\overrightarrow{k_{1}}}^{\ast}(\tau)u_{-\overrightarrow{k_{2}}}^{\ast}(\tau)\frac{d}{d\tau}u_{-\overrightarrow{k_{3}}}^{\ast}(\tau)+5perms]\nonumber\\
\label{con5}&&-12i\frac{(\mu-1)^{2}\Sigma}{\mu^{4}H^{4}}u_{\overrightarrow{k_{1}}}(0)u_{\overrightarrow{k_{2}}}(0)u_{\overrightarrow{k_{3}}}(0)\int_{-\infty}^{0}\frac{1}{\tau^{2}}d\tau(2\pi)^{3}\delta^{3}(\sum_{i}\overrightarrow{k_{i}})[\frac{\overrightarrow{k_{2}}\cdot\overrightarrow{k_{3}}}{k_{3}^{2}}\frac{d}{d\tau}u_{-\overrightarrow{k_{1}}}^{\ast}(\tau)u_{-\overrightarrow{k_{2}}}^{\ast}(\tau)\frac{d}{d\tau}u_{-\overrightarrow{k_{3}}}^{\ast}(\tau)+5perms]+c.c.\eea

In all these contributions above, we can substitute the explicit
forms of $\frac{d}{d\tau}u_{\overrightarrow{k}}^{\ast}$ into the
equation above to get a lot of integrals with $\tau$. It's
straightforward but the result is rather boring and page-wasting, so
we would put them into {\bf Appendix B}. Actually, it has 9 terms
differing from each an order of $\tau$ sequently, plus permutations
and complex conjugates. At the end we will compile all these terms
coming from contributions of all the terms in the Hamiltonian
according to their indices in a clearer form in order to make our
study comfortable.
\subsection{the shapes of bispectrum}
From Eqs. (\ref{con1}-\ref{con5}) and also
(\ref{con1'}-\ref{con5'}), we can integrate them out to have
different shapes. Since the power-law indices of $\tau$ in each
integration depend on the value of $\mu$, one may worry that for
indices less than $-1$, the infrared (IR) divergence will occur.
However, that is not the case. As has already been shown in Sec. III
B, the mode solution is frozen outside of horizon and there will be
no more IR evolution. Actually, the IR divergences will all be
canceled with each other and we do not see any real singularity,
though it is tedious to check analytically and even difficult
numerically. The same argument can be found in the paper JCAP {\bf
1004}, 027 (2010) as cited in \cite{Creminelli:2003iq}.\footnote{We
also thank Prof. Xingang Chen to remind this for us via email.}
Furthermore, we can replace the value of $u_{\overrightarrow{k}}$ at
far future $\tau=0$ to that at horizon-crossing time. Taking these
into consideration, we first write down all the possible shapes of
the bispectrum. Here we define \be
<|\zeta(\tau,\overrightarrow{k_1})\zeta(\tau,\overrightarrow{k_2})\zeta(\tau,\overrightarrow{k_3})|>=(2\pi)^{3}\delta^{3}(\sum_{i}\overrightarrow{k_{i}})\mathcal{B}(k_{1},k_{2},k_{3})\ee
where $\mathcal{B}(k_{1},k_{2},k_{3})$ are the shapes of
non-Gaussianity. There are totally 10 shapes at the leading order:
\bea \mathcal{B}_{3\mu-3}&=&\frac{(2\pi)^{\frac{3}{2}}\Sigma
H^{2}\cos(3\mu\pi)\Gamma(3\mu-2)}{8^{2}f_{R0}^{3}(-1)c_{s}^{6\mu-6}(k_{1}k_{2}k_{3})^{\mu+2}K^{3\mu-2}|\epsilon_{h}+\mu-1|^{3}\Gamma^{3}(-\mu+\frac{1}{2})}\nonumber\\
&&\label{b3mu-3}\left((1+\mu)(3\mu^{2}+\frac{\mu^{2}(1+\mu)}{2c_{s}^{2}}-12(\mu-1)^{2})\sum_{i>j}k_{i}^{2}k_{j}^{2}-(1+\mu)(\frac{\mu^{2}(1+\mu)}{4c_{s}^{2}}-6(\mu-1)^{2})\sum_{i}k_{i}^{4}\right)~,\eea
\bea \mathcal{B}_{3\mu-2}&=&\frac{(2\pi)^{\frac{3}{2}}\Sigma
H^{2}\cos(3\mu\pi)\Gamma(3\mu-1)}{8^{2}f_{R0}^{3}(-1)c_{s}^{6\mu-6}(k_{1}k_{2}k_{3})^{\mu+1}K^{3\mu-1}|\epsilon_{h}+\mu-1|^{3}\Gamma^{3}(-\mu+\frac{1}{2})}\nonumber\\
&&\left(\frac{1}{\mu}(\frac{\mu^{2}(1+\mu)}{2c_{s}^{2}}-6(\mu-1)^{2})\left(\prod_{i}\frac{1}{k_{i}}\right)\sum_{i\neq
j}(k_{i}^{2}k_{j}^{3}-k_{i}k_{j}^{4})+\frac{1}{\mu}(6\mu^{2}+\frac{\mu^{2}(1+\mu)}{c_{s}^{2}}-12(\mu-1)^{2})\sum_{i>j}k_{i}k_{j}\right)~,\eea

\bea
\mathcal{B}_{3\mu-1}&=&\frac{2(2\pi)^{\frac{3}{2}}H^{2}\cos(3\mu\pi)\Gamma(3\mu)}{8^{2}f_{R0}^{3}(-1)c_{s}^{6\mu-6}(k_{1}k_{2}k_{3})^{\mu}K^{3\mu}|\epsilon_{h}+\mu-1|^{3}\Gamma^{3}(-\mu+\frac{1}{2})}\left(\frac{\Sigma}{2c_{s}^{2}}\left(\prod_{i}\frac{1}{k_{i}}\right)(\sum_{i\neq
j}k_{i}k_{j}^{2}-\sum_{i}k_{i}^{3})-3(\Sigma+2\lambda)\right)~,\eea

\bea
\mathcal{B}_{\mu-3}&=&\frac{(2\pi)^{\frac{3}{2}}\mu^{5}(\mu-1)(1+\mu)^{3}H^{4}\cos(2\mu\pi)\Gamma(\mu-2)}{8^{3}c_{s}^{4\mu-2}f_{R0}^{2}(-1)(\prod_{i}k_{i}^{\mu+2})K^{\mu-2}|\epsilon_{h}+\mu-1|^{3}\Gamma^{3}(-\mu+\frac{1}{2})}\sum_{i}k_{i}^{2}~,\eea

\bea
\mathcal{B}_{\mu-2}&=&-\frac{2(2\pi)^{\frac{3}{2}}\mu^{4}(\mu-1)(1+\mu)^{2}H^{4}\cos(2\mu\pi)\Gamma(\mu-1)}{8^{3}c_{s}^{4\mu-2}f_{R0}^{2}(-1)(\prod_{i}k_{i}^{\mu+2})K^{\mu-2}|\epsilon_{h}+\mu-1|^{3}\Gamma^{3}(-\mu+\frac{1}{2})}\sum_{i}k_{i}^{2}~,\eea

\bea
\mathcal{B}_{\mu-1}&=&\frac{(2\pi)^{\frac{3}{2}}(\mu^{2}-1)H^{4}\cos(2\mu\pi)\Gamma(\mu)}{8^{2}f_{R0}^{2}(-1)c_{s}^{4\mu-4}(k_{1}k_{2}k_{3})^{\mu+2}K^{\mu}|\epsilon_{h}+\mu-1|^{3}\Gamma^{3}(-\mu+\frac{1}{2})}\nonumber\\
&&\left(9(\mu-1)(2(\mu-1)^{2}-\mu^{2})\sum_{i>j}k_{i}^{2}k_{j}^{2}-9(\mu-1)^{3}\sum_{i}k_{i}^{4}+\frac{\mu^{3}}{2c_{s}^{2}}\left(\sum_{i>j}k_{i}k_{j}\right)\sum_{i}k_{i}^{2}\right)~,\eea

\bea
\mathcal{B}_{\mu}&=&\frac{2(2\pi)^{\frac{3}{2}}H^{4}\cos(2\mu\pi)\Gamma(\mu+1)}{8^{2}f_{R0}^{2}(-1)c_{s}^{4\mu-4}(k_{1}k_{2}k_{3})^{\mu+2}K^{\mu+1}|\epsilon_{h}+\mu-1|^{3}\Gamma^{3}(-\mu+\frac{1}{2})}\nonumber\\
&&\left((\mu-1)\left(\prod_{i}k_{i}\right)(9(\frac{1}{\mu}-1)(2\mu-1)\sum_{i>j}k_{i}k_{j}+\frac{\mu^{2}}{2}\sum_{i}k_{i}^{2})+\frac{9(\mu-1)^{4}}{2\mu}\sum_{i\neq
j}(k_{i}^{2}k_{j}^{3}-k_{i}k_{j}^{4})\right)~,\eea

\bea
\mathcal{B}_{\mu+1}&=&\frac{18(2\pi)^{\frac{3}{2}}(\mu-1)^{2}H^{4}\cos(2\mu\pi)\Gamma(\mu+2)}{8^{3}f_{R0}^{2}(-1)c_{s}^{4\mu-4}(k_{1}k_{2}k_{3})^{\mu}K^{\mu+2}|\epsilon_{h}+\mu-1|^{3}\Gamma^{3}(-\mu+\frac{1}{2})}~,\eea

\bea
\mathcal{B}_{5\mu-5}&=&\frac{(2\pi)^{\frac{3}{2}}\Sigma^{2}(1+\mu)\cos(4\mu\pi)\Gamma(5\mu-4)}{8^{2}f_{R0}^{4}(-1)c_{s}^{8\mu-8}(k_{1}k_{2}k_{3})^{\mu+2}K^{5\mu-4}|\epsilon_{h}+\mu-1|^{3}\Gamma^{3}(-\mu+\frac{1}{2})}\left(2\sum_{i>j}k_{i}^{2}k_{j}^{2}-\sum_{i}k_{i}^{4}\right)~,\eea

\bea\label{b5mu-4}
\mathcal{B}_{5\mu-4}&=&\frac{2(2\pi)^{\frac{3}{2}}\Sigma^{2}\cos(4\mu\pi)\Gamma(5\mu-3)}{8^{2}f_{R0}^{4}(-1)c_{s}^{8\mu-8}(k_{1}k_{2}k_{3})^{\mu+2}K^{5\mu-3}\mu|\epsilon_{h}+\mu-1|^{3}\Gamma^{3}(-\mu+\frac{1}{2})}\left(\frac{1}{2}\sum_{i\neq
j}k_{i}^{2}k_{j}^{3}+\left(\prod_{i}k_{i}\right)\sum_{i>j}k_{i}k_{j}-\frac{1}{2}\sum_{i\neq
j}k_{i}k_{j}^{4}\right)~.\eea

For the next step, we have to discard the terms that will be
divergent as the indices in the original integral becomes less than
$-1$, and sum up all the convergent terms to give the total shape.
It depends on the value of $\mu$, obviously. In the following we
give the total shape for different values of $\mu$. The final result
is \be \mathcal{B}_{total}=\sum_y\Theta(y+1)\mathcal{B}_y~,\ee where
$\Theta$ is the Heaviside step function defined as: \bea \Theta(x)=\left\{ \begin{array}{c} 0~,~~~{\rm for}~x<0~,\\
\\
1~,~~~{\rm for}~x>0,\\
\end{array}\right. \eea and $y$ for all the subscripts of $\mathcal{B}$
above. From this result, we can also divide all the shapes to four
classes:\\
i) $\mathcal{B}_{3\mu-1}$, $\mathcal{B}_{\mu-1}$,
$\mathcal{B}_{\mu}$, $\mathcal{B}_{\mu+1}$: Since we assume that the
slow-roll parameters $\epsilon_h$ and $\epsilon_\theta$ larger than
0, thus the parameter $\mu>0$, and the indices of these shapes are
larger than $-1$. So they will definitely contribute to
$\mathcal{B}_{total}$;\\
ii) $\mathcal{B}_{3\mu-2}$, $\mathcal{B}_{3\mu-3}$,
$\mathcal{B}_{5\mu-4}$, $\mathcal{B}_{5\mu-5}$: If $\mu>1$, i.e.,
the power spectrum gets a red index, the indices of these shapes are
larger than $-1$ and they will contribute to
$\mathcal{B}_{total}$;\\
iii) $\mathcal{B}_{\mu-3}$: If $\mu<1$, i.e., the power spectrum
gets a blue index, the indices of these shapes are smaller than $-1$
and
they will not contribute to $\mathcal{B}_{total}$; and\\
iv) $\mathcal{B}_{\mu-2}$: The indices of these shapes are at
``divide" values. So when $\mu>1$, they will contribute to
$\mathcal{B}_{total}$ while when $\mu<1$, they will not.

We can also define the estimator through \be
\mathcal{B}(k_{1},k_{2},k_{3})=\frac{6}{5}F_{NL}\{\frac{2\pi^{2}}{k_{1}^{3}}\frac{2\pi^{2}}{k_{2}^{3}}\mathcal{P}_{k_{1}}^{\zeta}\mathcal{P}_{k_{2}}^{\zeta}+2perms.\}~,\ee
so, \bea
F_{NL}&=&\frac{5}{6}\frac{\mathcal{B}(k_{1},k_{2},k_{3})}{\{\frac{2\pi^{2}}{k_{1}^{3}}\frac{2\pi^{2}}{k_{2}^{3}}\mathcal{P}_{k_{1}}^{\zeta}\mathcal{P}_{k_{2}}^{\zeta}+2perms.\}}\nonumber\\
\label{fnl}&=&\frac{40f_{R0}^{2}(-1)c_{s}^{4\mu-2}|\epsilon_{h}+\mu-1|^{2}\Gamma^{4}(-\mu+\frac{1}{2})}{3\pi^{2}\mu^{2}H^{4}\sum_{i>j}(k_{i}k_{j})^{-(2\mu+1)}}\mathcal{B}(k_{1},k_{2},k_{3})~\eea
for each shape listed above. The result is rather obvious by just
substituting each shape into Eq. (\ref{fnl}), so we will not list
them here in order to save the page. We will only show the
equilateral limits of these $F_{NL}$'s, of which $k_1=k_2=k_3=k$, in
the next paragraph.
\subsection{The Equilateral Limit ($k_1=k_2=k_3=k$)}

In this section, we can take the equilateral limit, namely,
$k_1=k_2=k_3=k$, which gives simpler form of $F_{NL}$. Following the
last paragraph, we have: \be\label{f3mu-3}
(F_{NL})_{3\mu-3}^{equil}=k^{2-2\mu}\frac{5(2\pi)^{\frac{3}{2}}\Sigma\cos(3\mu\pi)\Gamma(3\mu-2)\Gamma(-\mu+\frac{1}{2})}{24\times3^{3\mu-2}\pi^{2}f_{R0}(-1)\mu^{2}c_{s}^{2\mu-4}H^{2}|\epsilon_{h}+\mu-1|}(1+\mu)(3\mu^{2}+\frac{\mu^{2}(1+\mu)}{4c_{s}^{2}}-6(\mu-1)^{2})~,\ee
\be
(F_{NL})_{3\mu-2}^{equil}=k^{2-2\mu}\frac{5(2\pi)^{\frac{3}{2}}\Sigma\cos(3\mu\pi)\Gamma(3\mu-1)\Gamma(-\mu+\frac{1}{2})}{24\times3^{3\mu-1}\pi^{2}f_{R0}(-1)\mu^{3}c_{s}^{2\mu-4}H^{2}|\epsilon_{h}+\mu-1|}(6\mu^{2}+\frac{\mu^{2}(1+\mu)}{c_{s}^{2}}-12(\mu-1)^{2})~,\ee
\be
(F_{NL})_{3\mu-1}^{equil}=k^{2-2\mu}\frac{5(2\pi)^{\frac{3}{2}}\cos(3\mu\pi)\Gamma(3\mu)\Gamma(-\mu+\frac{1}{2})}{12\times3^{3\mu}\pi^{2}f_{R0}(-1)\mu^{2}H^{2}c_{s}^{2\mu-4}|\epsilon_{h}+\mu-1|}(\Sigma(\frac{1}{2c_{s}^{2}}-1)-2\lambda)~,\ee
\be
(F_{NL})_{\mu-3}^{equil}=\frac{5(2\pi)^{\frac{3}{2}}\mu^{3}(\mu-1)(1+\mu)^{3}\cos(2\mu\pi)\Gamma(\mu-2)\Gamma(-\mu+\frac{1}{2})}{192\times3^{\mu-2}\pi^{2}|\epsilon_{h}+\mu-1|}~,\ee
\be
(F_{NL})_{\mu-2}^{equil}=-\frac{5(2\pi)^{\frac{3}{2}}\mu^{2}(\mu-1)(1+\mu)^{2}\cos(2\mu\pi)\Gamma(\mu-1)\Gamma(-\mu+\frac{1}{2})}{96\times3^{\mu-2}\pi^{2}|\epsilon_{h}+\mu-1|}~,\ee
\be
(F_{NL})_{\mu-1}^{equil}=\frac{5(2\pi)^{\frac{3}{2}}(\mu^{2}-1)c_{s}^{2}\cos(2\mu\pi)\Gamma(\mu)\Gamma(-\mu+\frac{1}{2})}{8\times3^{\mu}\pi^{2}\mu^{2}|\epsilon_{h}+\mu-1|}(3(\mu-1)^{3}-3(\mu-1)\mu^{2}+\frac{\mu^{3}}{2c_{s}^{2}})~,\ee
\be
(F_{NL})_{\mu}^{equil}=\frac{5(2\pi)^{\frac{3}{2}}c_{s}^{2}\cos(2\mu\pi)\Gamma(\mu+1)\Gamma(-\mu+\frac{1}{2})}{24\times3^{\mu+1}\pi^{2}\mu^{3}|\epsilon_{h}+\mu-1|}(\mu-1)(\mu^{3}-18(\mu-1)(2\mu-1))~,\ee
\be
(F_{NL})_{\mu+1}^{equil}=\frac{5(2\pi)^{\frac{3}{2}}(\mu-1)^{2}c_{s}^{2}\cos(2\mu\pi)\Gamma(\mu+2)\Gamma(-\mu+\frac{1}{2})}{32\times3^{\mu+2}\pi^{2}\mu^{2}|\epsilon_{h}+\mu-1|}~,\ee
\be
(F_{NL})_{5\mu-5}^{equil}=k^{4-4\mu}\frac{5(2\pi)^{\frac{3}{2}}\Sigma^{2}(1+\mu)\cos(4\mu\pi)\Gamma(5\mu-4)\Gamma(-\mu+\frac{1}{2})}{24\times3^{5\mu-4}\pi^{2}f_{R0}^{2}(-1)\mu^{2}H^{4}c_{s}^{4\mu-6}|\epsilon_{h}+\mu-1|}~,\ee
\be\label{f5mu-4}
(F_{NL})_{5\mu-4}^{equil}=k^{4-4\mu}\frac{5(2\pi)^{\frac{3}{2}}\Sigma^{2}\cos(4\mu\pi)\Gamma(5\mu-3)\Gamma(-\mu+\frac{1}{2})}{12\times3^{5\mu-3}\pi^{2}f_{R0}^{2}(-1)\mu^{3}H^{4}c_{s}^{4\mu-6}|\epsilon_{h}+\mu-1|}~,\ee
and same as the shape, we have for total non-linear parameter: \be
(F_{NL})_{total}^{equil}=\sum_y\Theta(y+1)(F_{NL})_y^{equil}~.\ee

From the result above we can see that $(F_{NL})^{equil}$ has a
slight running behavior due to the deviation of $\mu$ from 1, that
is, due to the non-minimal coupling behavior. This is different from
the usual minimal coupling case which was studied in
\cite{Acquaviva:2002ud} where the equilateral limit of $F_{NL}$ is
independent of $k$. This is another result in this paper and will
later be confirmed with numerical calculations. One can also obtain
$F_{NL}$ in the local limit ($k_1\approx k_2\gg k_3$) and folded
limit ($k_1=2k_2=2k_3$). Whichever limits they are in, all the
values of $F_{NL}$ with different indices can also be divided into
four classes by the same criteria used for
$\mathcal{B}(k_1,k_2,k_3)$.

\subsection{An explicit example: non-minimal coupled chaotic inflation}
In order to support our long analytical derivation, we in this
section focus on an explicit model of non-minimal coupling
inflation. For simplicity but without losing generality, we consider
chaotic inflation, of which the potential has a quadratic form as
$V(\phi)=\lambda\phi^4/4$ where $\lambda$ is the coupling
coefficients. Furthermore, we set the non-minimal coupling term
$f(R,\phi)=\frac{R}{8\pi G}+\xi R\phi^2$ where $G$ is the Newtonian
gravitational constant and $\xi$ is the non-minimal coupling
coefficient. This model is indeed very interesting since with the
presence of non-minimal coupling term, the coefficient $\lambda$
doesn't need to go to incredibly small value (Fakir and Unruh,
\cite{Salopek:1988qh}) to meet the observational constraint, and
this property has been used to construct Higgs inflation models,
which connects inflation theory to particle physics in the Standard
Model \cite{CervantesCota:1995tz}. In Higgs inflation, the Higgs
potential asymptotically coincides with the chaotic potential in
inflation period where the scalar field is in a high energy region
with some large value, and a too small value of $\lambda$ will make
the models inconsistent with the constraints from Standard Model
\cite{Bezrukov:2007ep}. Putting aside its physical motivation, in
this paper we study whether it can give rise to large
non-Gaussianities.

From the original action (\ref{action}), one can obtain the equation
of motion for the fields: \be \ddot\phi+3H\dot\phi-6\xi(\dot
H+2H^2)\phi+\frac{\partial V(\phi)}{\partial\phi}=0~, \ee and
Friedmann equation: \be 3H^2(\frac{1}{8\pi
G}+\xi\phi^2)=\frac{1}{2}\dot\phi^2+V(\phi)-6\xi H\phi\dot\phi~, \ee
and the above two equations can be combined to give another
equation: \be \dot H=(\frac{1}{8\pi
G}+\xi\phi^2)=-\frac{1}{2}\dot\phi^2+\xi
H\phi\dot\phi-\xi\dot\phi^2-\xi\phi\ddot\phi~.\ee

\begin{figure}[htbp]
\includegraphics[scale=0.3]{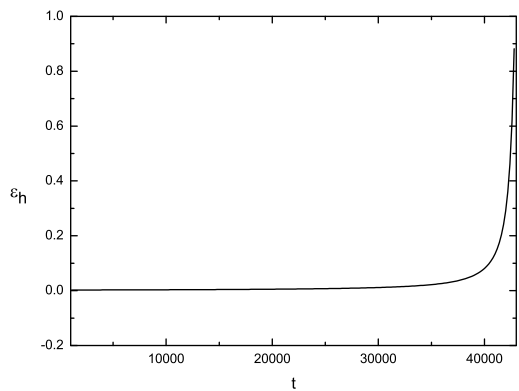}
\caption{The evolution of slow roll parameter $\epsilon$ w.r.t.
cosmic time $t$. The arrival of $\epsilon$ at 1 stops the inflation.
Parameters and initial values: $\xi=1000$, $\lambda=10^{-3}$,
$\phi_i=4.9$, $\dot\phi_i=0.063$. The normalization is $8\pi
G=1$.}\label{epsilon}
\end{figure}

\begin{figure}[htbp]
\includegraphics[scale=0.3]{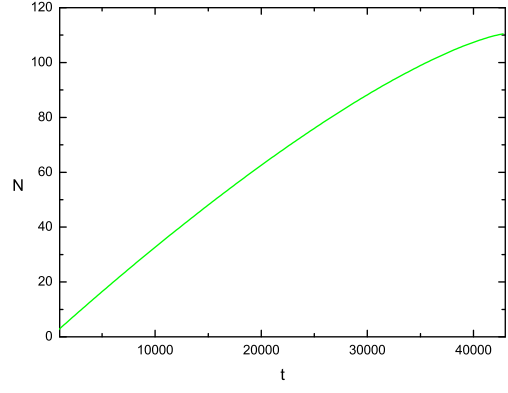}
\caption{The e-folding number $N$ w.r.t. cosmic time $t$. Parameters
and initial values are the same as in Fig.
\ref{epsilon}.}\label{efold}
\end{figure}

\begin{figure}[htbp]
\includegraphics[scale=0.3]{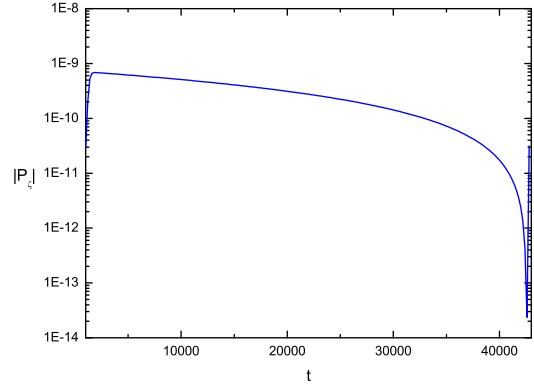}
\caption{The amplitude of power spectrum w.r.t. cosmic time
$t$.}\label{spectrumplot}
\end{figure}

\begin{figure}[htbp]
\includegraphics[scale=0.3]{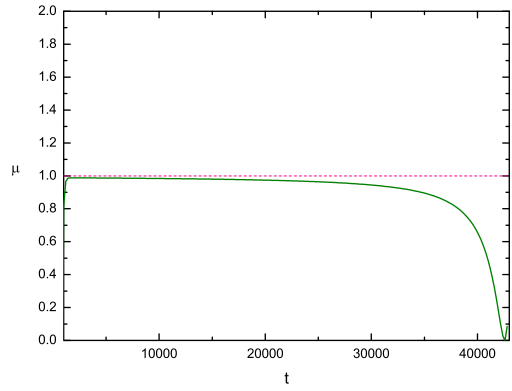}
\caption{$\mu$ w.r.t. cosmic time $t$. Slightly deviation from 1 is
obtained due to the non-minimal coupling.}\label{mu}
\end{figure}

Firstly, we draw the background evolution of the system in Figs.
\ref{epsilon} and \ref{efold}. From the plots we can see that with
natural choice of initial conditions and parameters, a period of
inflation can be easily constructed with enough amount of number of
e-folds. Setting $\xi=1000$, the parameter $\lambda$ could be raised
up to ${\cal O}(10^{-3})$, compared to the unnatural choice of
$\lambda\sim 10^{-14}$ in the case with $\xi=0$ (Fakir and Unruh,
\cite{Salopek:1988qh}). Figs. \ref{spectrumplot} and \ref{mu} shows
the amplitude and the $k-$dependence of its quadratic perturbation
spectrum, whose analytical form has already been given in Eq.
(\ref{spectrum}). For the given initial conditions and parameters,
we can see that the spectrum behaves nearly $k$-independent, with a
slight tilt caused by the deviation of $\mu$ from 1, which is mildly
favored by the WMAP-7 data \cite{Larson:2010gs}. The amplitude of
the spectrum is also consistent with the observations.

\begin{figure}[htbp]
\includegraphics[scale=0.4]{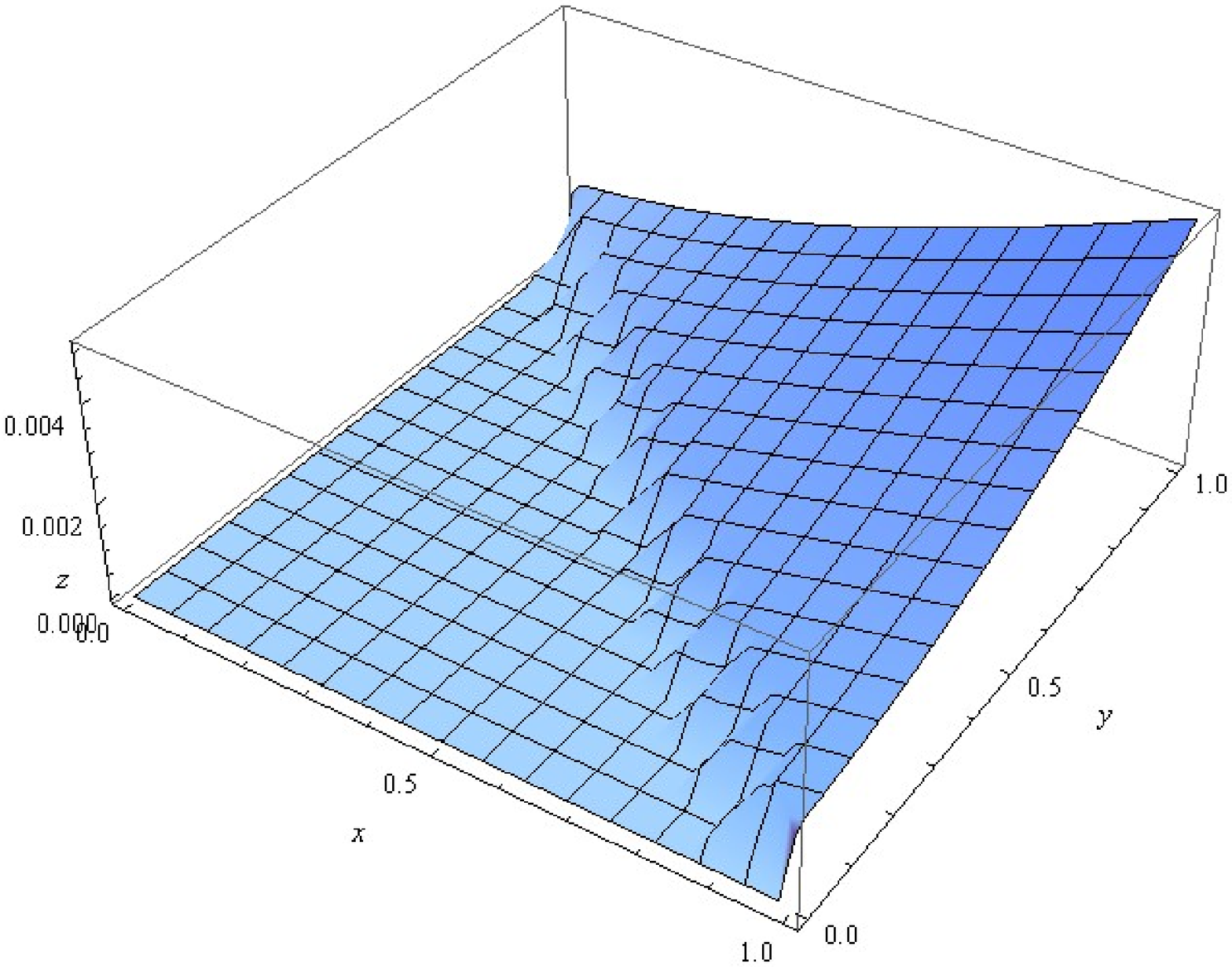}
\caption{The shape of non-Gaussianities
$\mathcal{B}(k_{1},k_{2},k_{3})$. Here we renormalize $x\equiv
k_1/k_3$, $y\equiv k_2/k_3$, and set $k_3=1$.}\label{shape}
\end{figure}

Next let's move on to the non-Gaussianities that this model can give
rise to. Since the power spectrum of this model has a blue tilt, the
first two classes of the total four in the shape
$\mathcal{B}(k_{1},k_{2},k_{3})$ as well as the estimator $F_{NL}$
which was shown in the last paragraphs will be applied. From our
numerical calculations, we can obtain the values of every parameter
that appears in Eqs. (\ref{b3mu-3})-(\ref{b5mu-4}) as well as
(\ref{f3mu-3})-(\ref{f5mu-4}). With this in hand, we can easily
obtain the numerical results of the non-Gaussianities generated in
this model. The total shape (of leading order) of the
non-Gaussianities and the estimator in the equilateral limit
$(F_{NL})^{equil}$ are shown in Figs. \ref{shape} and \ref{fnl}.
From the plots we can see that the shape of the non-Gaussianities
are well within the constraints of the observational data. Note that
the estimator $(F_{NL})^{equil}$ shows a running behavior with
respect to $k(=k_1=k_2=k_3)$, with a positive sign. This is because
of the effect of non-minimal coupling which makes the parameter
$\mu$ deviate from $1$ and thus $(F_{NL})^{equil}$ will be dependent
on $k$. In the GR limit $\mu\rightarrow 1$, $(F_{NL})^{equil}$ will
have a constant value, as shown in \cite{Acquaviva:2002ud}.

\begin{figure}[htbp]
\includegraphics[scale=0.6]{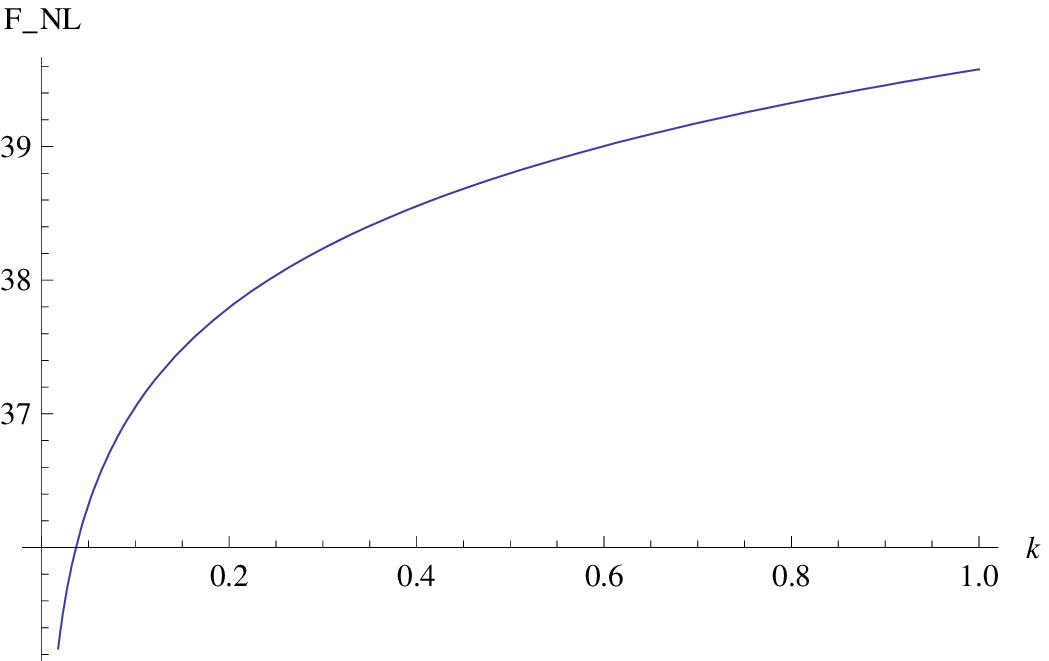}
\caption{The estimator of non-Gaussianities in the equilateral
limit: $(F_{NL})^{equil}$. Running behavior is obtained due to the
non-minimal coupling effect.}\label{fnl}
\end{figure}
\section{conclusion and discussions}

In this paper, we performed the non-Gaussianities of a general
single scalar field which linearly couples to gravity. Our result
shows that due to the non-minimal coupling, the power spectrum will
deviate from scale-invariance, which in order lead to the
complicated non-Gaussianities in the 3rd order. We obtained all the
possible shapes of the 3-point correlation functions and for
different tilt of power spectrum, we showed that different shapes
will be involved in to give rise to non-Gaussianities. Our
calculation presents the description in general non-minimal coupling
inflation and this result, if verified to all the orders, can
provide relation between 2- and 3-point correlation functions and
can be used to constrain non-minimal coupling models.

Another result that was presented in this paper is that there is
some running behavior of the estimator $F_{NL}$ in the equilateral
limit with respect to $k$ which is different from the normal minimal
coupling case. This behavior is due to the non-minimal coupling, and
are expected to have signature on observations in order to
distinguish minimal and non-minimal cases.

Besides the analytical calculations, we also performed numerical
computations on a specific example of non-minimal coupling chaotic
inflation model. This model is extendedly studied with the
application of Higgs inflation. We obtained the behavior of
background, 2-point power spectrum as well as the shape and
estimator of its non-Gaussianity. We showed that the
non-Gaussianities are well within the observational constraint, with
the running behavior of $F_{NL}^{equil}$ w.r.t. $k$.

Other than inflation, such non-minimal coupling system can also be
applied to other aspect in cosmology. For example, non-minimal
coupling theory can act as dark energy \cite{Uzan:1999ch} or give
rise to a bouncing/cyclic universe \cite{Abreu:1994fd}. Moreover,
non-minimal coupling can be used to make up an open/closed universe
\cite{Linde:1995rv}, while dualities of Einstein's gravity in the
presence of a non-minimal coupling was taken on in
\cite{Abramo:2003cr}. The stabilities and singularities in
superacceleration phases was discussed in \cite{Gunzig:2000kk}, and
the removal of singularities in Loop Quantum Gravity with
non-minimal coupling was studied in \cite{Bojowald:2006bz}. Our
calculation of non-Gaussianities are also expected to be applied to
these interesting fields.
\section*{Acknowledgments}

TQ owes many thanks to Dr. Xian Gao, Dr. Yi Wang and Prof. Xingang
Chen for useful help and guidance. TQ also thanks Dr. Yi-Fu Cai and
Dr. Antonio de Felice for discussions. TQ is grateful to the High
Energy Accelerator Research Organization (KEK) and Institute of the
Physic and Mathematics of the Universe (IPMU) in Japan for their
hospitalities during his visit, and the ``Horiba International
Conference Cosmo/Cospa 2010" and the following string workshop held
in Tokyo University for stimulating talks and discussions when this
work was processing, as well as the LeCospa group meeting in
National Taiwan University for helpful discussions. This research is
supported in parts by the National Science Council of R.O.C. under
Grant No. NSC99-2112-M-033-005-MY3 and No. NSC99-2811-M-033-008 and
by the National Center for Theoretical Sciences.

\appendix
\section{ extend to non-linear coupling}
Here, we only list the constraint equations for the general form of
non-minimal coupling single scalar field, including the non-linear
coupling case. The constraints from $N$ and $N_i$ are: \bea
&&4\alpha\Sigma+12Hf_{R0}\dot{\zeta}-12H^{2}f_{R0}\alpha-4a^{-2}f_{R0}H\partial^{2}\psi\nonumber\\
&&-4a^{-2}f_{R0}\partial^{2}\zeta-(12\alpha H-6\dot{\zeta}+2a^{-2}\partial^{2}\psi)\partial_{t}f_{R0}+72(\dot{H}-H^{2})(\dot{H}+2H^{2})f_{RR0}\alpha-36(\dot{H}-H^{2})f_{RR0}\ddot{\zeta}\nonumber\\
&&-36(\dot{H}-H^{2})Hf_{RR0}(4\dot{\zeta}-\dot{\alpha})+12a^{-2}(\dot{H}-H^{2})f_{RR0}(\partial^{2}\dot{\psi}+2H\partial^{2}\psi+\partial^{2}\alpha+2\partial^{2}\zeta)\nonumber\\
&&+6H\partial_{t}[-12(\dot{H}+2H^{2})f_{RR0}\alpha+6Hf_{RR0}(4\dot{\zeta}-\dot{\alpha})+6f_{RR0}\ddot{\zeta}-2a^{-2}f_{RR0}(\partial^{2}\dot{\psi}+2H\partial^{2}\psi+\partial^{2}\alpha+2\partial^{2}\zeta)]\nonumber\\
&&+24a^{-2}(\dot{H}+2H^{2})f_{RR0}\partial^{2}\alpha-12a^{-2}Hf_{RR0}\partial^{2}(4\dot{\zeta}-\dot{\alpha})-12a^{-2}f_{RR0}\partial^{2}\ddot{\zeta}\nonumber\\
&&+4a^{-4}f_{RR0}\partial^{2}(\partial^{2}\dot{\psi}+2H\partial^{2}\psi+\partial^{2}\alpha+2\partial^{2}\zeta)=0\eea
and \bea
&&-\frac{1}{2}a^{-2}f_{R0}\partial^{2}\tilde{N}^{i}-12a^{-2}H(\dot{H}+2H^{2})f_{RR0}\partial^{i}\alpha+6a^{-2}H^{2}f_{RR0}\partial^{i}(4\dot{\zeta}-\dot{\alpha})+6a^{-2}Hf_{RR0}\partial^{i}\ddot{\zeta}\nonumber\\
&&-2a^{-4}Hf_{RR0}\partial^{i}(\partial^{2}\dot{\psi}+2H\partial^{2}\psi+\partial^{2}\alpha+2\partial^{2}\zeta)+a^{-2}\partial_{t}[12(\dot{H}+2H^{2})f_{RR0}\partial^{i}\alpha-6Hf_{RR0}\partial^{i}(4\dot{\zeta}-\dot{\alpha})\nonumber\\
&&-6f_{RR0}\partial^{i}\ddot{\zeta}+2a^{-2}f_{RR0}\partial^{i}(\partial^{2}\dot{\psi}+2H\partial^{2}\psi+\partial^{2}\alpha+2\partial^{2}\zeta)]\nonumber\\
&&+2a^{-2}f_{R0}H\partial^{i}\alpha+a^{-2}\partial_{t}f_{R0}\partial^{i}\alpha-2a^{-2}f_{R0}\partial^{i}\dot{\zeta}=0\eea
respectively, where $f_{R0}$ and $f_{RR0}$ denotes the background
value of the first and second derivatives of $f(R,\phi)$ with
respect to $R$.
\section{Contributions from terms in $H_{int}^p$ w.r.t. $\tau$}
Here, we list the contributions from terms in $H_{int}^p$, i.e. Eq.
(\ref{con1}-\ref{con5}), in terms of $\tau$ by substituting
$\frac{d}{d\tau}u_{\overrightarrow{k}}^{\ast}$ in. One can see that
there contains integrals of different power-laws of $\tau$, each
differing one order from the other in every term. One can combine
the integrals of the same order power-law to have more neat forms,
as in {\bf Sec. III D}.

The contribution from $\dot{\zeta}^{3}$:  \bea <\dot{\zeta}^{3}>
&\supset&-6\frac{(2\pi)^{3}\delta^{3}(\sum_{i}\overrightarrow{k_{i}})(\Sigma+2\lambda)\pi^{\frac{3}{2}}H^{2}e^{-3i\frac{\mu\pi}{2}}}{8^{\frac{3}{2}}f_{R0}^{3}(-1)c_{s}^{3\mu-6}(k_{1}k_{2}k_{3})^{\mu}|\epsilon_{h}+\mu-1|^{3}\Gamma^{3}(-\mu+\frac{1}{2})}\int_{-\infty}^{0}d\tau
e^{ic_{s}K\tau}|\tau|^{3\mu-1}\nonumber\\
\label{con1'}&&-18\frac{(2\pi)^{3}\delta^{3}(\sum_{i}\overrightarrow{k_{i}})(\mu-1)^{2}\pi^{\frac{3}{2}}H^{4}e^{-3i\frac{\mu\pi}{2}}}{8^{\frac{3}{2}}f_{R0}^{2}(-1)c_{s}^{3\mu-6}(k_{1}k_{2}k_{3})^{\mu}|\epsilon_{h}+\mu-1|^{3}\Gamma^{3}(-\mu+\frac{1}{2})}\int_{-\infty}^{0}d\tau
e^{ic_{s}K\tau}|\tau|^{\mu+1}+c.c.\eea

The contribution from $\zeta\dot{\zeta}^{2}$: \bea
<\zeta\dot{\zeta}^{2}>&\supset&-6\frac{(2\pi)^{3}\delta^{3}(\sum_{i}\overrightarrow{k_{i}})\Sigma\pi^{\frac{3}{2}}\mu
H^{2}e^{-3i\frac{\mu\pi}{2}}}{8^{\frac{5}{2}}f_{R0}^{3}(-1)c_{s}^{3\mu}(k_{1}k_{2}k_{3})^{\mu+2}|\epsilon_{h}+\mu-1|^{3}\Gamma^{3}(-\mu+\frac{1}{2})}\nonumber\\
&&\times\int_{-\infty}^{0}d\tau
e^{ic_{s}K\tau}[\{+4c_{s}^{4}\mu(1+\mu)k_{2}^{2}k_{3}^{2}|\tau|^{3\mu-3}+8ic_{s}^{5}k_{1}k_{2}^{2}k_{3}^{2}|\tau|^{3\mu-2}\}+2perms.]\nonumber\\
&&-18\frac{(2\pi)^{3}\delta^{3}(\sum_{i}\overrightarrow{k_{i}})\pi^{\frac{3}{2}}\mu(\mu-1)^{2}H^{4}e^{-3i\frac{\mu\pi}{2}}}{8^{\frac{5}{2}}f_{R0}^{2}(-1)c_{s}^{3\mu}(k_{1}k_{2}k_{3})^{\mu+2}|\epsilon_{h}+\mu-1|^{3}\Gamma^{3}(-\mu+\frac{1}{2})}\nonumber\\
\label{con2'}&&\times\int_{-\infty}^{0}d\tau
e^{ic_{s}K\tau}[\{+4c_{s}^{4}\mu(1+\mu)k_{2}^{2}k_{3}^{2}|\tau|^{\mu-1}+8ic_{s}^{5}k_{1}k_{2}^{2}k_{3}^{2}|\tau|^{\mu}\}+2perms.]+c.c.\eea

The contribution from $\zeta(\partial\zeta)^{2}$: \bea
<\zeta(\partial\zeta)^{2}>&\supset&-\frac{2(2\pi)^{\frac{9}{2}}\delta^{3}(\sum_{i}\overrightarrow{k_{i}})(\mu-1)\mu^{2}H^{4}e^{-3i\frac{\mu\pi}{2}}}{8^{3}c_{s}^{3\mu}f_{R0}^{2}(-1)(\prod_{i}k_{i}^{\mu+2})|\epsilon_{h}+\mu-1|^{3}\Gamma^{3}(-\mu+\frac{1}{2})}\int_{-\infty}^{0}d\tau[(\overrightarrow{k_{2}}\cdot\overrightarrow{k_{3}})e^{ic_{s}K\tau}\{\mu^{3}(1+\mu)^{3}|\tau|^{\mu-3}\nonumber\\
\label{con3'}&&+2ic_{s}\mu^{2}(1+\mu)^{2}K|\tau|^{\mu-2}-4c_{s}^{2}\mu(1+\mu)\left(\sum_{i>j}k_{i}k_{j}\right)|\tau|^{\mu-1}-8ic_{s}^{3}(k_{1}k_{2}k_{3})|\tau|^{\mu}\}+2perms.]+c.c.\eea

The contribution from $\dot{\zeta}(\partial\zeta)^{2}$: \bea
<\dot{\zeta}(\partial\zeta)^{2}>&\supset&-\frac{12(2\pi)^{3}\delta^{3}(\sum_{i}\overrightarrow{k_{i}})\pi^{\frac{3}{2}}(\mu-1)^{2}H^{4}e^{-3i\frac{\mu\pi}{2}}}{8^{\frac{5}{2}}f_{R0}^{2}(-1)c_{s}^{3\mu-2}(k_{1}k_{2}k_{3})^{\mu+2}|\epsilon_{h}+\mu-1|^{3}\Gamma^{3}(-\mu+\frac{1}{2})}\nonumber\\
&&\times\int_{-\infty}^{0}d\tau[(\overrightarrow{k_{2}}\cdot\overrightarrow{k_{3}})k_{1}^{2}e^{ic_{s}K\tau}[\mu^{2}(1+\mu)^{2}|\tau|^{\mu-1}+2ic_{s}\mu(1+\mu)(k_{2}+k_{3})|\tau|^{\mu}-4c_{s}^{2}k_{2}k_{3}|\tau|^{\mu+1}]+2perms]\nonumber\\
&&-\frac{4(2\pi)^{3}\delta^{3}(\sum_{i}\overrightarrow{k_{i}})\pi^{\frac{3}{2}}\Sigma
H^{2}e^{-3i\frac{\mu\pi}{2}}}{8^{\frac{5}{2}}f_{R0}^{3}(-1)c_{s}^{3\mu-2}(k_{1}k_{2}k_{3})^{\mu+2}|\epsilon_{h}+\mu-1|^{3}\Gamma^{3}(-\mu+\frac{1}{2})}\nonumber\\
&&\times\int_{-\infty}^{0}d\tau[(\overrightarrow{k_{2}}\cdot\overrightarrow{k_{3}})k_{1}^{2}e^{ic_{s}K\tau}[\mu^{2}(1+\mu)^{2}|\tau|^{3\mu-3}+2ic_{s}\mu(1+\mu)(k_{2}+k_{3})|\tau|^{3\mu-2}-4c_{s}^{2}k_{2}k_{3}|\tau|^{3\mu-1}]+2perms]\nonumber\\
\label{con4'}&&+c.c.\eea

The contribution from $\dot{\zeta}\partial\zeta\partial\chi$: \bea
<\dot{\zeta}\partial\zeta\partial\chi>&\supset&\frac{18(2\pi)^{3}\delta^{3}(\sum_{i}\overrightarrow{k_{i}})\pi^{\frac{3}{2}}(\mu-1)^{4}H^{4}e^{-3i\frac{\mu\pi}{2}}}{8^{\frac{5}{2}}f_{R0}^{2}(-1)c_{s}^{3\mu}(k_{1}k_{2}k_{3})^{\mu+2}\mu|\epsilon_{h}+\mu-1|^{3}\Gamma^{3}(-\mu+\frac{1}{2})}\nonumber\\
&&\times\int_{-\infty}^{0}d\tau
e^{ic_{s}K\tau}[\frac{\overrightarrow{k_{2}}\cdot\overrightarrow{k_{3}}}{k_{3}^{2}}\{+4c_{s}^{4}\mu(1+\mu)k_{1}^{2}k_{3}^{2}|\tau|^{\mu-1}+8ic_{s}^{5}k_{2}k_{1}^{2}k_{3}^{2}|\tau|^{\mu}\}+5perms]\nonumber\\
&&+\frac{2(2\pi)^{3}\delta^{3}(\sum_{i}\overrightarrow{k_{i}})\pi^{\frac{3}{2}}\Sigma^{2}e^{-3i\frac{\mu\pi}{2}}}{8^{\frac{5}{2}}f_{R0}^{4}(-1)c_{s}^{3\mu}(k_{1}k_{2}k_{3})^{\mu+2}\mu|\epsilon_{h}+\mu-1|^{3}\Gamma^{3}(-\mu+\frac{1}{2})}\nonumber\\
&&\times\int_{-\infty}^{0}d\tau
e^{ic_{s}K\tau}[\frac{\overrightarrow{k_{2}}\cdot\overrightarrow{k_{3}}}{k_{3}^{2}}\{+4c_{s}^{4}\mu(1+\mu)k_{1}^{2}k_{3}^{2}|\tau|^{5\mu-5}+8ic_{s}^{5}k_{2}k_{1}^{2}k_{3}^{2}|\tau|^{5\mu-4}\}+5perms]\nonumber\\
&&+\frac{12(2\pi)^{3}\delta^{3}(\sum_{i}\overrightarrow{k_{i}})\pi^{\frac{3}{2}}(\mu-1)^{2}\Sigma
H^{2}e^{-3i\frac{\mu\pi}{2}}}{8^{\frac{5}{2}}f_{R0}^{3}(-1)c_{s}^{3\mu}(k_{1}k_{2}k_{3})^{\mu+2}\mu|\epsilon_{h}+\mu-1|^{3}\Gamma^{3}(-\mu+\frac{1}{2})}\nonumber\\
\label{con5'}&&\times\int_{-\infty}^{0}d\tau
e^{ic_{s}K\tau}[\frac{\overrightarrow{k_{2}}\cdot\overrightarrow{k_{3}}}{k_{3}^{2}}\{+4c_{s}^{4}\mu(1+\mu)k_{1}^{2}k_{3}^{2}|\tau|^{3\mu-3}+8ic_{s}^{5}k_{2}k_{1}^{2}k_{3}^{2}|\tau|^{3\mu-2}\}+5perms]+c.c.\eea

\end{document}